\documentstyle[aps,preprint]{revtex}

\input epsf.sty

\begin{document}

\tightenlines

\title{
Financial multifractality and its subtleties: an example of DAX
}

\author{A.Z. G\'orski$^1$\thanks{Corresponding author. Tel.:
+48-12-662 8240; fax: +48-12+662 8458. {\it E--mail address:}
Andrzej.Gorski@ifj.edu.pl .},
S. Dro\.zd\.z$^{1-4}$, J. Speth$^3$}

\address{$^1$Institute of Nuclear Physics, PL--31-342 Krak\'ow, Poland \\
$^2$Institute of Physics, University of Rzesz\'ow,
PL--35-310 Rzesz\'ow, Poland \\
$^3$Institut f\"ur Kernphysik, Forschungszentrum J\"ulich,
D--52425 J\"ulich, Germany \\
$^4$Physikalisches Institut, Universit\"at Bonn, D-53115 Bonn, Germany}

\date{May 17, 2002}

\maketitle

\begin{abstract}
Detailed study of multifractal characteristics of the financial time
series of asset values and of its returns is performed
using a collection of the high frequency Deutsche Aktienindex data.
The tail index ($\alpha$), the Renyi exponents based on the box
counting algorithm for the graph
($d_q$) and the generalized Hurst exponents ($H_q$) are computed
in parallel for short and daily return times.
The results indicate a more complicated nature of the stock market
dynamics than just consistent multifractal.

\end{abstract}

\section{Introduction}

Comprising the financial time series into a unique description
on different time scales by means
of the multifractal formalism is an attractive perspective
both for fundamental as well as for practical reasons~\cite{Mandel}.
As a constructive, though also critical, contribution towards
this aim in the present paper we systematically review the related basic
characteristic of the financial time series
using the large collection of the high-frequency
Deutsche Aktienindex (DAX) data with over $10^6$ data records
within the time period 28.~Nov~1997 $\div$ 30.~Dec~1999
\cite{Data1},
as well as the daily data since 29.~Sep~1959 $\div$
4.~Sep~2001 \cite{Data2} (over $10^4$ data points).
We thus estimate the tail index for the distribution of DAX
returns, the Renyi exponents for the DAX index and return's graphs,
as well as the corresponding generalized Hurst exponents.
It should be stressed that contrary to all the previous related studies
the fractal exponents for the graphs are here
computed directly, using the (2--dimensional) box counting
algorithm, independently of the Hurst exponent computations.

 Let $S(t)$ represent an index value as a function of time.
The rate of return within the time lag $\tau$ (the return time)
is usually defined as
\begin{equation}
G'_{\tau}(t) = {S(t+\tau) - S(t) \over S(t)} \ .
\label{Gprim}
\end{equation}
The logarithmic returns
\begin{equation}
G_{\tau}(t) = \ln S(t+\tau) - \ln S(t) =
\ln \left( 1+G'_{\tau}(t) \right)
\simeq G'_{\tau}(t)
\end{equation}
are more often used when analyzing the financial time series.
They are additive with respect to the return time,
\begin{equation}
G_{\tau_1}(t) + G_{\tau_2}(t+\tau_1) = G_{\tau_1+\tau_2}(t)
\ .
\end{equation}
 This implies translational symmetry of returns and scaling invariance
of the asset price (return is identical for prices in any currency).
 The logarithmic returns are indistinguishable from normal returns
if $\vert G'_{\tau}(t)\vert \ll 1$.
 In general, the normal returns (\ref{Gprim}) are slightly bigger in
absolute
value which leads to fatter tails due to the inequality $x > \ln (1+x)$
for $x\in(0, 1)$.
Simple index differences have also been
investigated but they are dimensional quantities \cite{normtails}.
 Throughout this paper by the index $x_t$ at time $t$ we will understand
the logarithm of its value,
$x_t \ = \ \ln S_t$.
Hence, the returns $G_{\tau}$ are increments of the index $x_t$,
just as the white noise (WN) random variables are increments of the
Brownian random walk (BRW).
The index $x_t$ can be viewed as integrated returns
\begin{equation}
x_{(t_2+1)\tau} \ = \ x_{t_1\tau} + \sum_{k=t_1}^{t_2} G_\tau(k\tau)
\ ,
\label{increments}
\end{equation}
where $x_{t_1\tau}$ is the integration constant.
 To compare various returns the normalized returns are introduced
that have zero average and unit variance
\begin{equation}
g_{\tau}(t) = { G_{\tau}(t) - \langle G_{\tau}(t) \rangle_t
\over \sigma(G_{\tau}(t)) }
\ ,
\label{normrets}
\end{equation}
where $\langle \dots \rangle_t$ denotes averaging over time
variable and $\sigma(\dots)$ is the standard deviation.

 In the following section the probability density function (pdf)
of the DAX returns is discussed in order to demonstrate that it
shares essentially all the global characteristics identified so far
in the literature for other indices, though some small quantitative
differences do of course take place.
 Section~III is devoted to study self similarity of the DAX index
and of its returns using the (2--dimensional) box counting algorithm.
 The generalized Hurst exponents are analyzed in Sec.~IV.
Summary and conclusions are given in the last section.

\section{Distribution of returns}

 The basic property of normalized returns that has been intensively
investigated in recent years is their pdf
(obtained by constructing histograms from the empirical data).
 The old assumption that returns follow the uncorrelated BRW
with independent identically distributed ({\it iid})
Gaussian random variables
\cite{Bachelier} has been challenged by Mandelbrot and by Fama
\cite{Mandel63,Fama63}. Mandelbrot analyzed a relatively short
(about 2000 data points) time series of cotton prices
and he reached conclusion that returns have L\'evy stable
symmetric distribution with Pareto (power like) fat tails
\cite{Levy,Pareto}.

 This result has been refined in recent years by the Boston
group \cite{Stanley95,Stanley99} using over $10^6$ high frequency
data points of the S$\&$P500 index.
They have found clear departures from the
L\'evy stable distribution. First, suggesting the exponential
truncation of fat tails for returns $> 6 \sigma$, where $\sigma$
is the standard deviation. Later, they found a power law fit with
the tail index exceeding the L\'evy distribution \cite{Stanley99}.
The last result was confirmed by other groups for
distributions of returns of other indices, including the
FOREX data \cite{Evertsz2}.

\begin{figure}[bht]
\centering
\epsfxsize=8.0truecm
\mbox{
\epsfbox{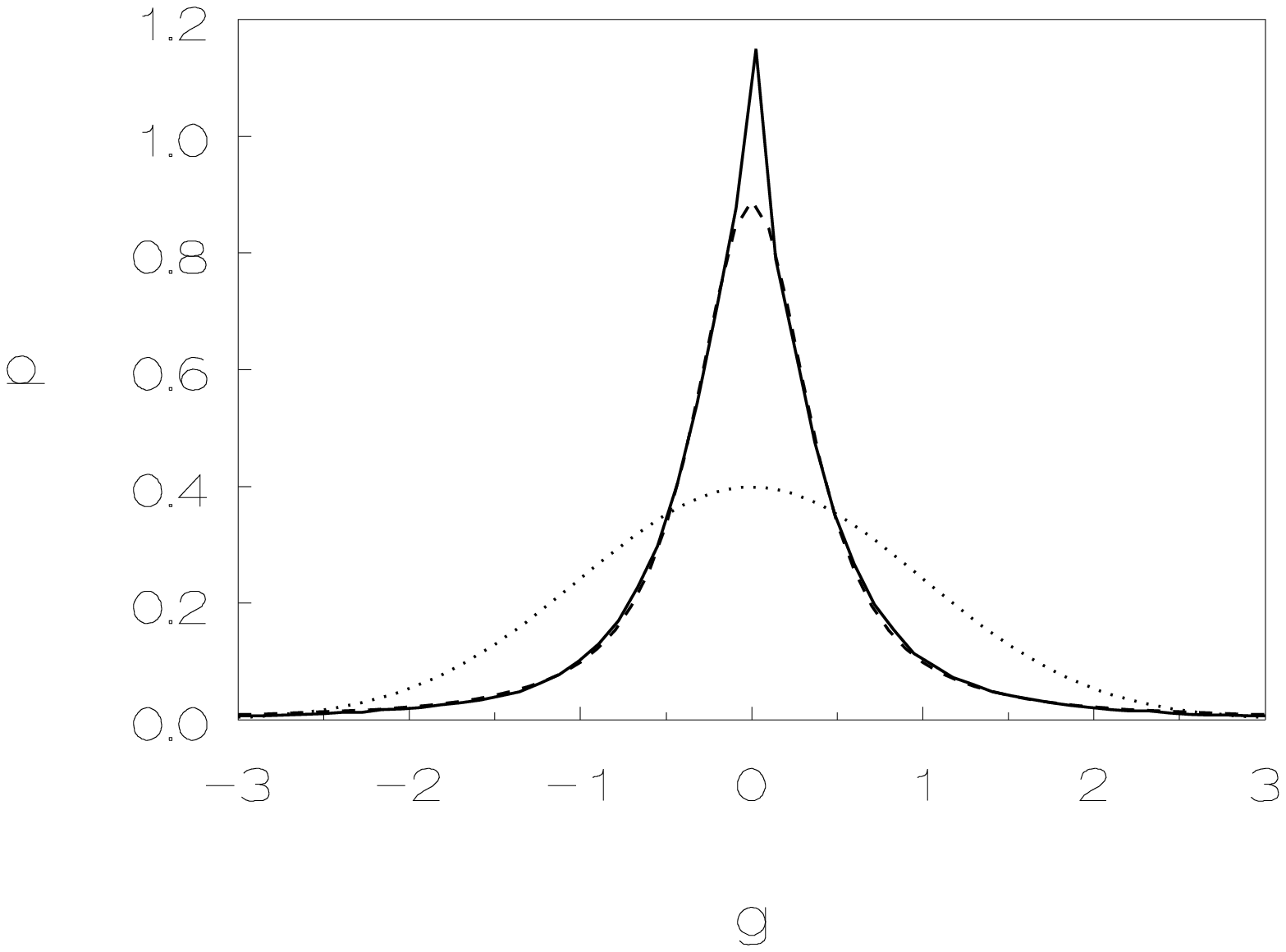}
}
\caption{Histogram for DAX normalized minutely returns. Normalized
Gauss (dotted line) and best fit for L\'evy stable distribution
(dashed line, $\alpha=1.16$ and $\gamma=0.286$)
are also displayed.
}
\label{fig:Fig1}
\end{figure}

 In this Section we investigate histograms of the DAX returns
for high and mid--frequency data.
 In Fig.~\ref{fig:Fig1} we plot the minutely data histogram
($\tau=1$~min.)
for normalized returns (\ref{normrets}). The normalized Gaussian
distribution (dotted line)
as well as the best fit for the L\'evy stable distribution (dashed
line) are also plotted.
 It is clear that the high frequency returns are neither
Gaussian nor L\'evy stable. The best fit for L\'evy distribution
within the central region $g\in [-g_I; +g_I]$, with $g_I\simeq 5$
gives for the L\'evy parameters $\alpha = 1.16$ and $\gamma = 0.286$.
It is interesting to notice
that this fit is very stable with respect to the range of the
variable $g$: one gets identical results within about 1\%
for $g_I=1, 3$ and $10$.
The tail index ($\alpha$) obtained in this way is too small
to fit well for $g>2$.
Also, it is smaller than the value 1.4 for S\&P500 reported in
\cite{Stanley99}.

 Here, one should notice one exception: the central bin contains
abundant number of data points (and it was not taken into
account for the fits).
This phenomenon we call the {\it zero return enhancement} and it
is typical for high frequency data, especially for
single stocks, indices that do not contain too many stocks
and for not too large stock markets.
For short time lags it happens that there are no transactions,
the index value is constant and the return distribution
is enhanced for the single return value $g=0$.
For $15$~sec. DAX returns this effects gives about factor 2
enhancement for the central bin and still considerable enhancement
for $\tau=1$~min.
However, for much larger S\&P500 index (and larger NY stock market)
this effect is hardly visible even for minutely data.
 Because of this effect and discreteness of
returns, we prefer the least square fit adopted for
L\'evy distribution \cite{levyfit}
within the central region
of the histogram, with the zero return bin neglected,
instead of using the scaling form of the return probability
$p(0)$ (see also \cite{Stanley95} and Sec.~IVA in \cite{Stanley99}).

 The log--log plot corresponding to Fig.~\ref{fig:Fig1}
is given in Fig.~\ref{fig:Fig5}(A).
For comparison the dashed line with slope corresponding to the tail
index $\alpha = 2.4$ is also plotted.
We can see that both, negative and positive returns give
approximately the same tail index close to the value given above.
This value considerably exceeds the range of the L\'evy stable
distributions. Our estimates give $\alpha \simeq 2.5$,
in particular
$\alpha=2.4$ for positive and $\alpha=2.6$ for negative returns.
 Here, one can observe an effect that is reverse to what was found in
other markets: the negative return tail has over 8\% bigger exponent
than the positive tail.
Quality of the linear fit for the $n$ data points with
coordinates $(x_i, y_i)$ is usually measured by the linear correlation
coefficient $r$ (also known as the Pearson's $r$),
where $r\in \langle -1, +1 \rangle$. For $r=1$ the fit is of course
perfect.
Here, in both cases the correlation coefficient $r > 0.996$
within the range $2 < g < 20$.
The resulting tail index, $\alpha$,
is considerably below the S\&P500 result ($\alpha\simeq 2.8$),
reported in \cite{Stanley99} (though, in both cases well
above the L\'evy stable limit $\alpha=2$).
In practical terms the DAX develops somewhat fatter tails
which reflects its more sizable fluctuations (more "rare events").

 For $g > 20$ (or $\log_2 g \ge 4.5$) the bins of the histogram
contain very few data points ($0\div3$ for reasonable bin width
that is $\simeq 0.1$ for Fig.~\ref{fig:Fig5}(A))
and any regression for those rare events is rather
doubtful. Hence, any fit for the very far tail
($g > 20$) truncation seems to be not well justified even for the
largest sets of the high frequency data available today.
 The same results have been also obtained for $\tau=15$~sec.
 In summary, the high frequency DAX return distribution appears
to be quite complicated consisting of the zero return enhancement
in bin with $g=0$, L\'evy like distribution in the central region
($\vert g\vert \le 2$) with $\alpha\simeq 1.3$,
inverse power like behavior with $\alpha\simeq 2.4$
for the tail ($2 < g < 20$) and not easily quantifable
behavior for very large events ($g > 20$).

 The above picture, with two different $\alpha$'s for the central
region and the tail, seems to be consistent with recent
suggestions that the return distribution is a union of different
distributions for returns in "normal" day time and "rush hours",
like soon after opening, before closing and at 14:30 in Europe
 \cite{Huang00,Stan1,Stan2,Hupp}.
 In this paper we however do not extract special subsets of our financial
time series that would screen out such effects as they constitute
an essential element of the stock market dynamics.
Doing so would also shrink the data set which even as a whole
seems to be of relatively modest size (see {\it e.g.}
\cite{Weron}).
Furthermore, in addition to trends in the special subsets mentioned
above there exist
some other trends on various time-scales which are not connected with
the calendar time but are associated with the internal market
dynamics (bubbles and crashes, see \cite{Johan,Stan3} for a recent
discussion).
Hence, throughout this paper we analyze the complete DAX data set
available.

\begin{figure}[bht]
\centering
\epsfxsize=8.0truecm
\mbox{
\epsfbox{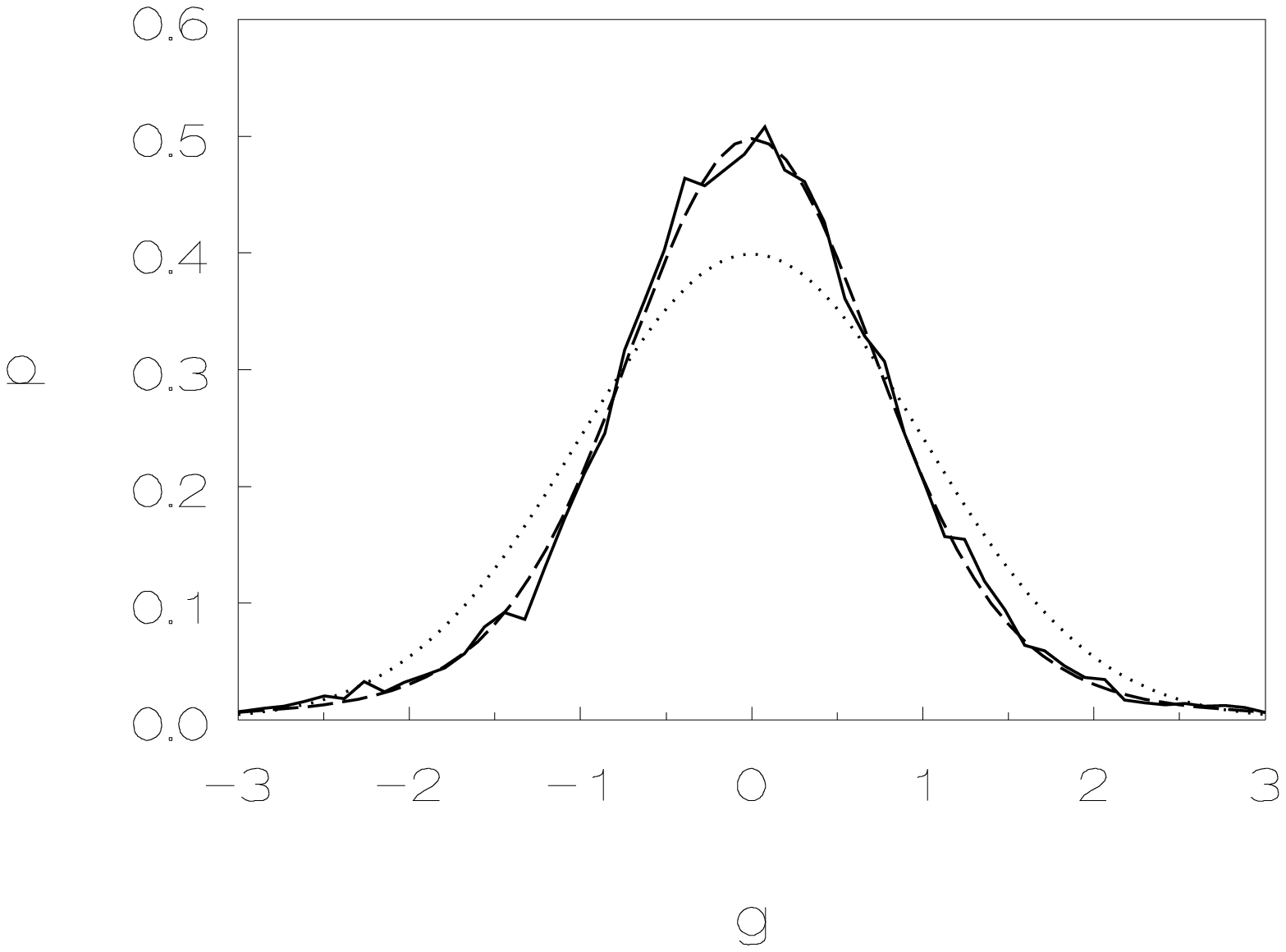}
}
\caption{Histogram for DAX normalized 10~min. returns. Normalized
Gauss (dotted line) and best fit for L\'evy stable distribution
(dashed line, $\alpha=1.44$ and $\gamma=0.327$)
are also displayed.
}
\label{fig:Fig2}
\end{figure}

\begin{figure}[bht]
\centering
\epsfxsize=8.0truecm
\mbox{
\epsfbox{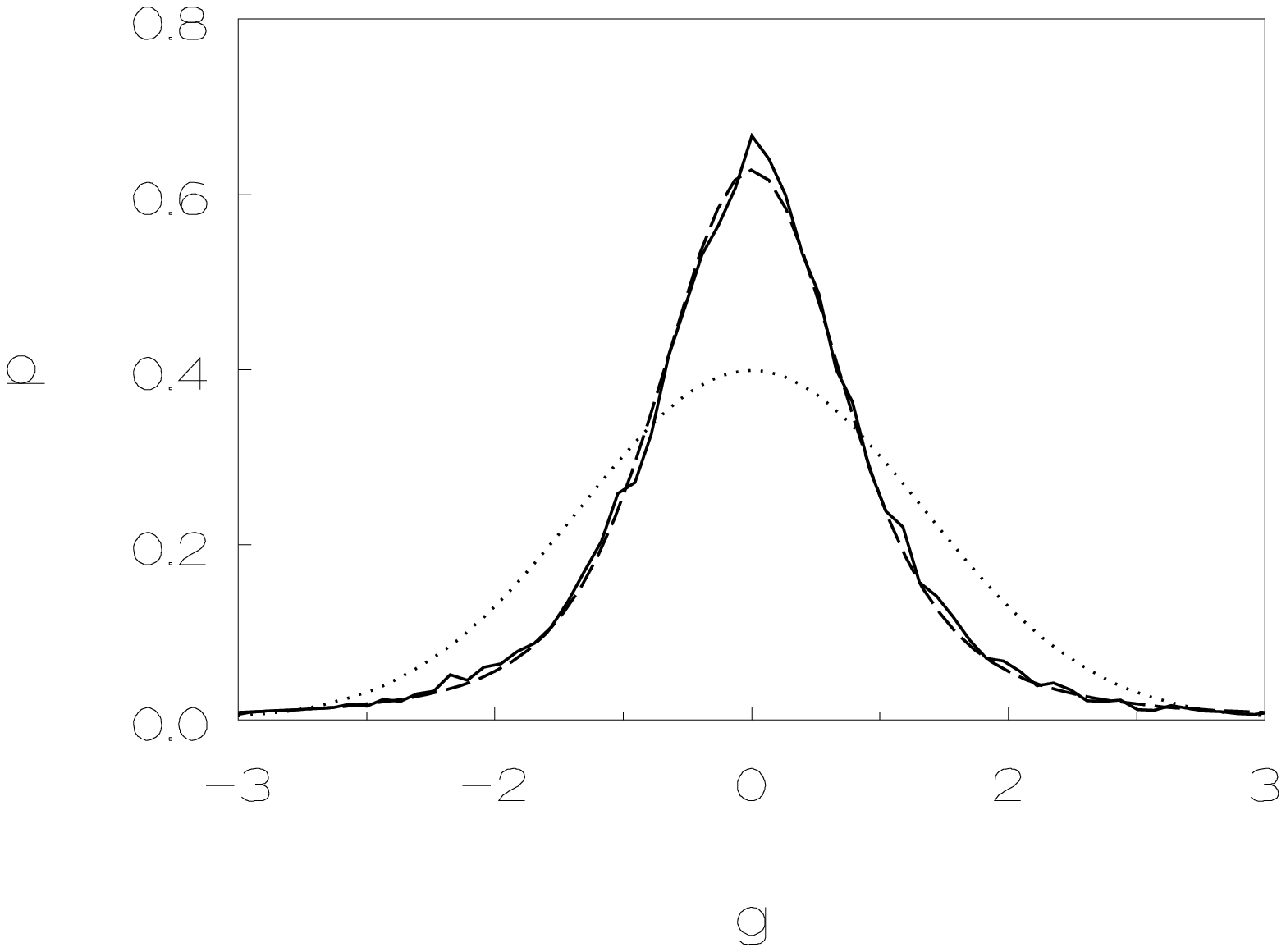}
}
\caption{Histogram for DAX normalized 1~hour returns. Normalized
Gauss (dotted line) and best fit for L\'evy stable distribution
(dashed line, $\alpha=1.40$ and $\gamma=0.354$)
are also displayed.
}
\label{fig:Fig3}
\end{figure}

 Increase in the return time, $\tau$, leads to the increase of
the parameter $\alpha$ of the fitted L\'evy curve.
In Figs.\ref{fig:Fig2},\ref{fig:Fig3} and
Figs.\ref{fig:Fig5}(B,C) the same plots as in
Figs.\ref{fig:Fig1},\ref{fig:Fig5}(A) are displayed but for
return times $\tau = 10$~min. and $\tau = 1$~hour, respectively.
 In both cases we have good fit of the L\'evy distribution
in the central region, $g\in (-g_I, +g_I)$, $g_I\simeq 3$.
The L\'evy parameters $\alpha$ and $\gamma$ slowly
grow from the values $\alpha\simeq 1.2$ and $\gamma\simeq0.29$
for the minutely returns up to $\alpha \simeq 1.4$ and
$\gamma\simeq0.39$ for the hourly returns
(see Figs.~\ref{fig:Fig1},\ref{fig:Fig2},\ref{fig:Fig3}).
At the same time the tail index outside the central region
grows from $\alpha=2.4$ for the minutely returns up to
$\alpha\simeq3.5$ for hourly returns
(see Figs.~\ref{fig:Fig5}(A--C)).

\begin{figure}[bht]
\centering
\epsfxsize=8.0truecm
\mbox{
\epsfbox{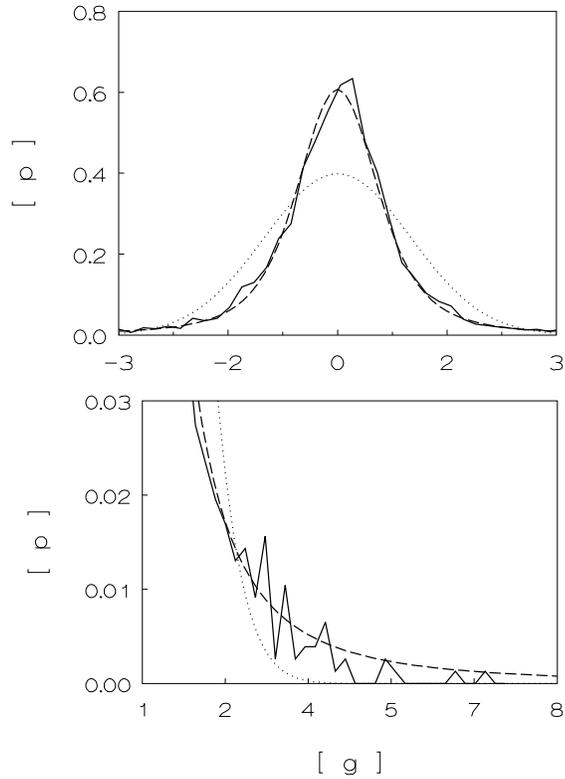}
}
\caption{Histogram for DAX normalized daily returns. Normalized
Gauss (dotted line) and best fit for L\'evy stable distribution
(dashed line, $\alpha=1.7$ and $\gamma=0.385$)
are also displayed. Lower panel: magnification
of the tail behavior.
}
\label{fig:Fig4}
\end{figure}

\begin{figure}[bht]
\centering
\epsfxsize=14.0truecm
\mbox{
\epsfbox{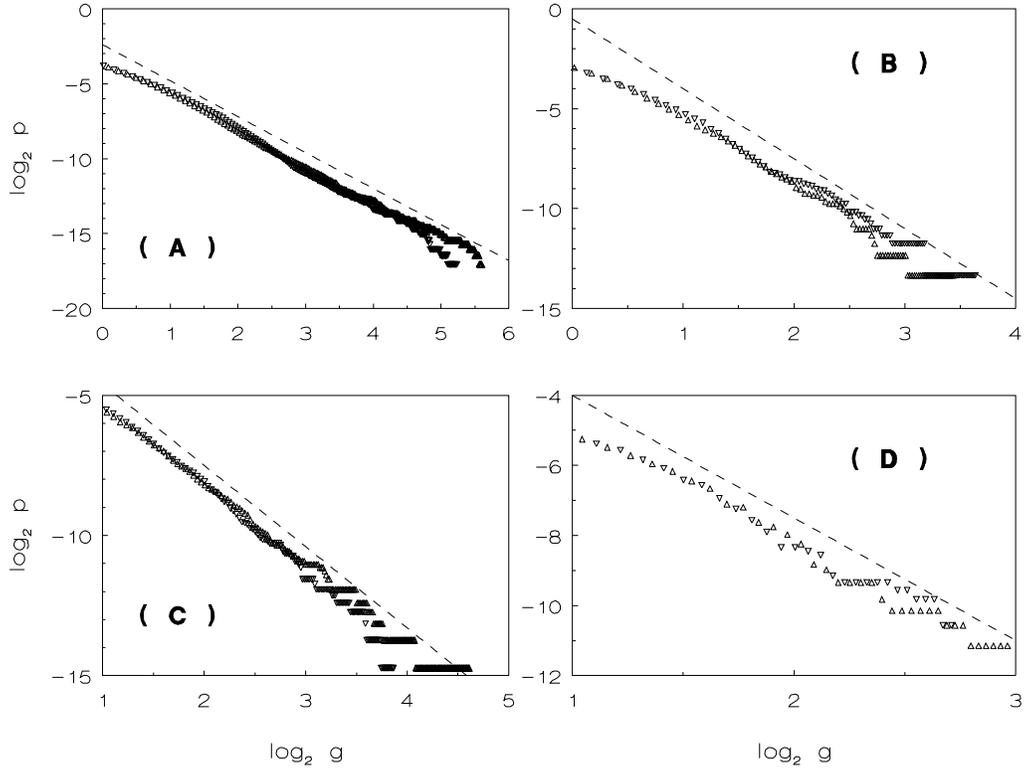}
}
\caption{Log--log plot for normalized minutely DAX returns
for return times $= 1$, $10$, $60$~min. and 1 day,
respectively for plots (A--D).
The triangles represent positive tail while the inverted
triangles, the negative tail.
The dashed line corresponds to the tail index $\alpha = 2.4$,
$2.9$, $3.5$ and $3.5$ for graphs (A--D).
}
\label{fig:Fig5}
\end{figure}

 For return times longer than $\tau=60$~min. the number of
data points is decreasing below $10^3$ which is not sufficient
to precisely estimate the tail behavior.
 Hence, for mid--frequency data we use much longer data set
\cite{Data2} with historical daily data to have reasonably
long time series ($>10^3$).
The return time here, $\tau \simeq 500$~min., is over 3 orders of
magnitude longer than for the shortest available
($\tau=15$~sec.) data.
Again, the central region where the histogram can be well
approximated by a L\'evy stable distribution is:
$\vert g \vert < g_I \simeq 3$. The best fit
for the L\'evy parameters in this region (also, quite stable
with respect of the changes in $g_I$) is:
$\alpha \simeq 1.7$ and $\gamma\simeq 0.385$
(see Fig.~\ref{fig:Fig4}).
In the lower panel, we have added 40--fold magnified
tail to show that even for the largest available set of data
the tail behavior cannot be reasonably estimated already for
$g>5$. For longer return times the situation is even worse.
For $\tau = 30$~min.
the tail index seems to be stabilized at the hourly return value
$\alpha\simeq3.5$
(Fig.~\ref{fig:Fig5}(D)).
In all cases it is greater than the maximum value
allowed for the L\'evy stable distributions ($\alpha < 2.0$),
in agreement with what was found for other indices.

 Good fit of the L\'evy stable distribution within the central region
is consistent with the old result obtained by Mandelbrot for
a data set of quite modest size \cite{Mandel63}.
 However, change of the L\'evy parameters with the return time
($\tau$), as well as differences in the value of the L\'evy parameter
$\alpha$ for the central region and for the tail in each case
clearly shows the complexity of the return distributions
and their incompatibility with a plain L\'evy curve.
 Also, from lower panel of
Fig.~\ref{fig:Fig4}
one can see apparent difference in the tail behavior of
the normalized return distributions and the normalized Gaussian
distribution.

 The behavior of rare events ({\it i.e.} for tail with
normalized returns $g > 20$) cannot be reasonably estimated
for return times $\tau>1$~min., at least using the standard
histogram building and fit method, due to the insufficient number
of such events in the present day financial data available.
Similarly as for the S$\&$P500 \cite{Stanley99}, even though
on short time scales the DAX return distributions are clearly
not L\'evy stable,
in the range from $\tau=1/4$~min. up to $\tau=1$~day
we do not observe convergence of the pdf to the Gaussian
(see Figs.~\ref{fig:Fig1}--\ref{fig:Fig4}).

\begin{figure}[bht]
\centering
\epsfxsize=8.0truecm
\mbox{
\epsfbox{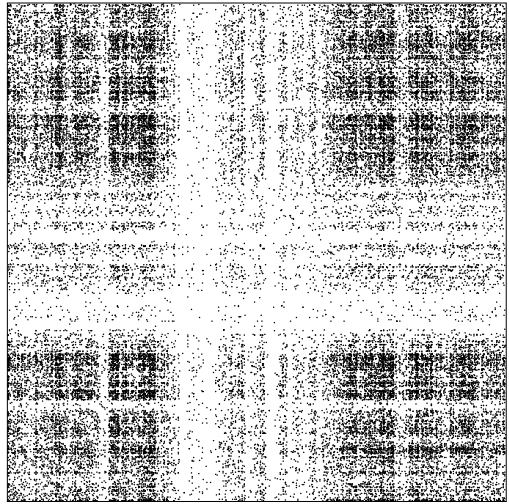}
}
\caption{Recurrence plot for DAX 10~min. returns. Time is on both axis,
the embedding dimension is set to $E = 5$ and the normalized neighborhood
size is $\varepsilon=0.2$.}
\label{fig:Fig6}
\end{figure}

For the standard (Lindeberg) version of the Central Limit Theorem
(CLT) the random variables must be {\it iid} and the stochastic
process must be stationary. This condition is not satisfied for financial
time series where, in particular, various short as well as long
range cycles are present.
 In particular, empirical data display clear nonstationarity as can
be seen from the pattern of the recurrence
plot that is given for 10~min. returns data series in
Fig.~\ref{fig:Fig6}. Here, the relation between closeness in time and
in phase space is displayed using the Takens phase space reconstruction
method. The embedding dimension is taken
$E = 5$, but its exact value is not important in this case as the recurrence
plot is not sensitive with respect to this parameter.
The normalized neighborhood size was set $\varepsilon=0.2$,
but similar structures were obtained for smaller $\varepsilon$
(see {\it e.g.} \cite{Schreiber} Sec.~III and
\cite{Eckmann,Schreiber98} for more details), as well as for
the embedding dimension up to $E=9$.
 Recently, Bouchaud {\it et al} \cite{Bouchaud} have shown that for
a special model with correlations similar as in financial series
a slow convergence to the Gaussian pdf can be proven and it is visible
for return times $\tau > 1$~day (see also \cite{Stanley99}).

\section{Self similarity}

 Distribution of returns gives us only a small part of the whole
information included in the time series. In particular,
the time order is completely neglected.
To account for the time dependence one should analyze
the whole (2--dimensional) graph of the functions $x_t = x(t)$ and
$g_{\tau}(t)$ ($D_G$ in the notation of \cite{Mandel}).
 Although, the projections of the data points at the time and
at the value axis are pseudofractal sets with zero
Hausdorff--Besicovitch dimension and their pseudofractal
scaling exponents are equal to one \cite{pseudofr},
the graph in principle can be a regular fractal as is the
case for the WN and BRW.
 The fractal properties of returns have been suggested long ago
by Mandelbrot \cite{MandelbrotII}. More detailed analysis
has been published by Evertsz {\it et al} \cite{Evertsz2,Evertsz1}.
However, it should be stressed that their analysis was based
on the calculation of the generalized Hurst exponents
(see Sec.~IV).

To investigate fractal properties of DAX index and its returns
we apply directly the standard box counting algorithm for computation
of the Renyi box counting exponents defined by \cite{Renyi}
\begin{equation}
d_q = {1\over 1-q} \ \lim_{N\to\infty} { \ln \sum_i p_i^q(N)
\over \ln N} \equiv \lim_{N\to\infty} {\ln Y(N) \over \ln N}
\ ,
\label{dqdef}
\end{equation}
where $N$ is the total number of "boxes" (bins), $p_i$
is the part of the "mass" ({\it i.e.} fraction of all points)
contained in the $i$-th box.

\begin{figure}[bht]
\centering
\epsfxsize=14.0truecm
\mbox{
\epsfbox{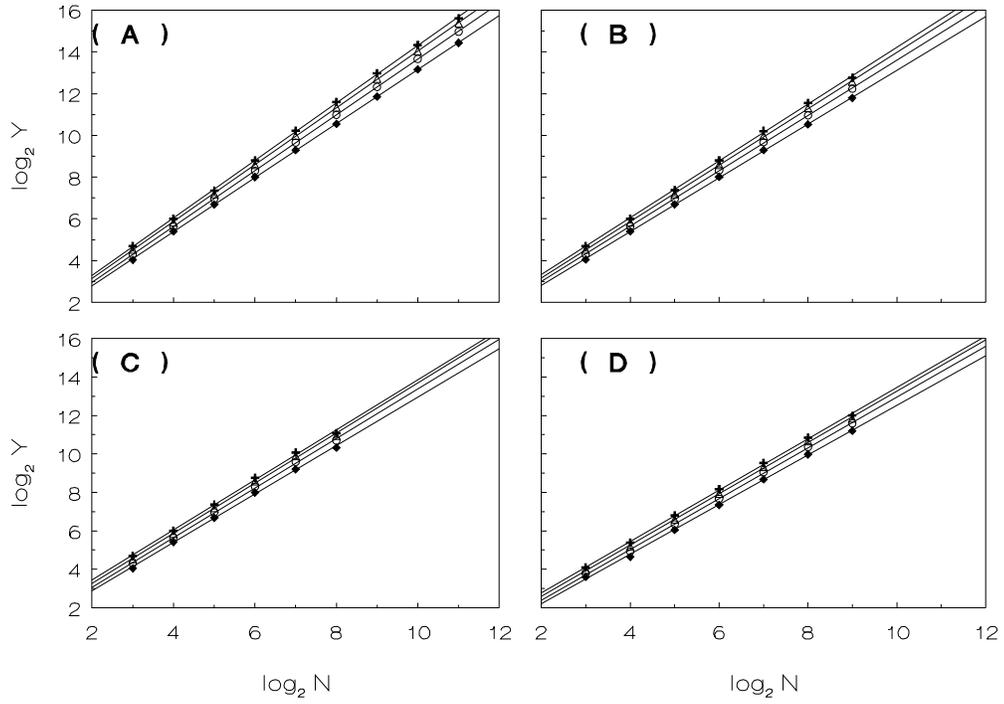}
}
\caption{Calculating Renyi box counting exponents, $d_q$, for log of DAX.
$q=0.5$, $1$, $2$, $4$ for crosses, triangles, circles,
and diamonds, respectively. Plots from A to D
are given for return times $\tau=1$, $10$, $60$~min. and
$1$~day. Solid lines represent least square linear fits.
}
\label{fig:Fig7}
\end{figure}

 Computation is performed for both graphs, $x(t)$ and $g_\tau(t)$.
To the best of our knowledge this is the first direct calculation
of the fractal dimensions of the graphs for any financial index and
its returns. Usually, the self similar properties were inferred
from the Hurst exponents or from the rescale range analysis,
see {\it e.g.} \cite{Skjeltrop,Ausloos01}.
As will be shown in the following sections these two methods give
different results, especially for the returns' graphs.
In Fig.~\ref{fig:Fig7} results of the calculations of $d_q$
exponents are displayed for the logarithm of DAX ($x_t$).
 The calculations for time lags
$\tau = 1, 10, 60$~min., 1 day and for
$q = 0.5, 1, 2$ and $4$ are shown.
 In fact, the same results were obtained for the plain
index ($S_t$).
This supports the view that the graph is a real fractal,
not just a pseudofractal as the exponents for fractals are
invariant with respect to the homeomorphic transformations.
In all cases one gets quite reasonable linear fits.
The correlation parameter ($r$), as can be expected, is lowest
for $\tau=60$~min, where the number of original data points is the
smallest (below $10^4$). However, even in this case we have
$r > 0.998$ (for all other cases we have $r > 0.9994$),
quite a reasonable value.
The linear scaling extends through $6\div9$ points (binary orders of
magnitude) at the log--log plot which is also a standard range
for numerical estimates of fractal dimensions.
 The resulting Renyi box counting exponents are given in Table~\ref{Tab1}.
Even though, the range of the return time ($\tau$) and the
range of the index $q$ are relatively large, all values for the
exponents $d_q$ are within the interval $d_q \in [ 1.3, 1.4 ]$,
well below the value characteristic for the BRW
(in all Tables the last digit is not the significant digit).
 This suggests that we have a self similar and close to monofractal
curve for the DAX index.
 In addition, one can observe a general tendency that the index
$d_q$ is slowly decreasing for longer time lags except
for $\tau=1$~day, where another set of data was used.

\begin{figure}[bht]
\centering
\epsfxsize=14.0truecm
\mbox{
\epsfbox{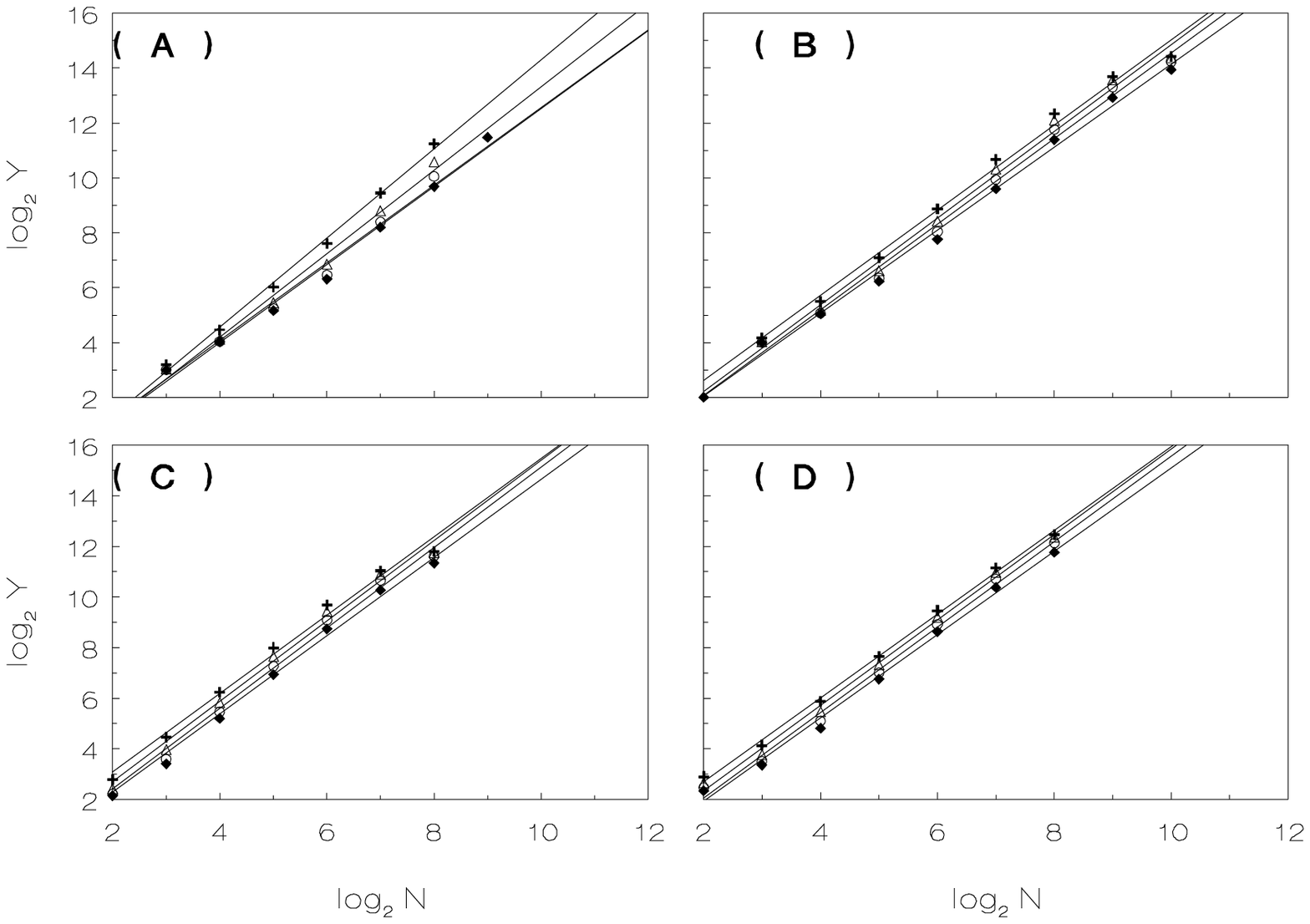}
}
\caption{Calculating Renyi box counting exponents for DAX returns.
$q=0.5$, $1$, $2$, $4$ for crosses, triangles, circles,
and diamonds, respectively. Plots from A to D
are given for return times $\tau=1$, $10$, $60$~min. and
$1$~day. Solid lines represent least square linear fits.
}
\label{fig:Fig8}
\end{figure}

Similar analysis applied to the normalized DAX returns
leads to the result displayed in Fig.~\ref{fig:Fig8}.
Here, the quality of the linear fit in the log--log plot is not
as convincing as in the previous case (Fig.~\ref{fig:Fig7})
of the index itself, especially for the smaller $q$-values.
Extracting nevertheless the corresponding linear best fit $(r > 0.994)$
coefficients results in the values for $d_q$ which are more dispersed
and spread within the range $d_q = 1.4\div 1.7$,
depending on $q$ as well as on the return time, $\tau$.

Our results thus indicate that, within a few percent, which is
comparable to the estimated numerical accuracy,
the $q$--dependence (and $\tau$--dependence) of $d_q$ is rather weak.
Hence, the corresponding graphs can be viewed as close to monofractal
and this indication is especially suggestive for the index
whose scaling properties are quite convincing. It is also worth to
notice that the corresponding $d_q$ values for the index
are systematically below the BRW value (1.5).
This may reflect an important element specific to the financial dynamics.
Somewhat less transparent is the related scaling behavior of the returns, 
though this behavior also carries some information about the underlying
dynamics and is therefore shown here.
Assuming however that some scaling approximately applies in this case as well,
it seems natural that the above estimated bounds limiting
variation of $d_q$ for returns point to values which are systematically
bigger than for the index, as the former is the differentiated index
(see eq.(\ref{increments})).
While this difference equals $0.5$ for WN and BRW, in our case
the difference is on average significantly smaller.
To avoid numerical artifacts we have repeated all calculations with
logarithm of the index and we have obtained the same results
within the numerical accuracy. This shows that the results are
invariant with respect to nonlinear transformations of the
data.
As the fractal dimensions and Renyi box counting exponents are related to
the Hurst exponents we will return to this issue in the
subsequent section.

\section{Generalized Hurst exponents}

 The Hurst exponent ($H_1$) has been introduced long ago in the
"rescale range theory" for measurements of the Nile flooding
and drought amplitudes \cite{Hurst}.
The generalized Hurst exponents, $H_q=H(q)$, for a time series
$g(t)$ ($t=1,2,\ldots$) are defined by
the scaling properties of its structure functions
$S_q(\tau)$ \cite{Mandel}
\begin{equation}
S_q(\tau) = \langle \vert g(t+\tau) - g(t) \vert^q
\rangle^{1/q}_T \sim \tau^{H(q)}
\label{HurstDef}
\end{equation}
where $q>0$, $\tau$ is the time lag and averaging is over the time
window $T\gg\tau$, usually the largest time scale of the system.
 The function $H(q)$ contains information about
averaged generalized volatilities at scale $\tau$ (only
$q=1,2$ are used to define the volatility).
In particular, the $H_1$ exponent indicates persistent ($H_1>1/2$)
or antipersistent ($H_1<1/2$) behavior of the trend.
For the BRW (brown noise, $1/f^2$)
one gets $H_q=1/2$, while for the pink ($1/f$) and WN we have
$H_q=0$.
 For the popular L\'evy stable and truncated L\'evy processes
with parameter $\alpha$ it has been found that
$H_q = q / \alpha$ for $q<\alpha$ and
$H_q=1$ for $q\ge\alpha$ \cite{Lovejoy}.

\begin{figure}[bht]
\centering
\epsfxsize=14.0truecm
\mbox{
\epsfbox{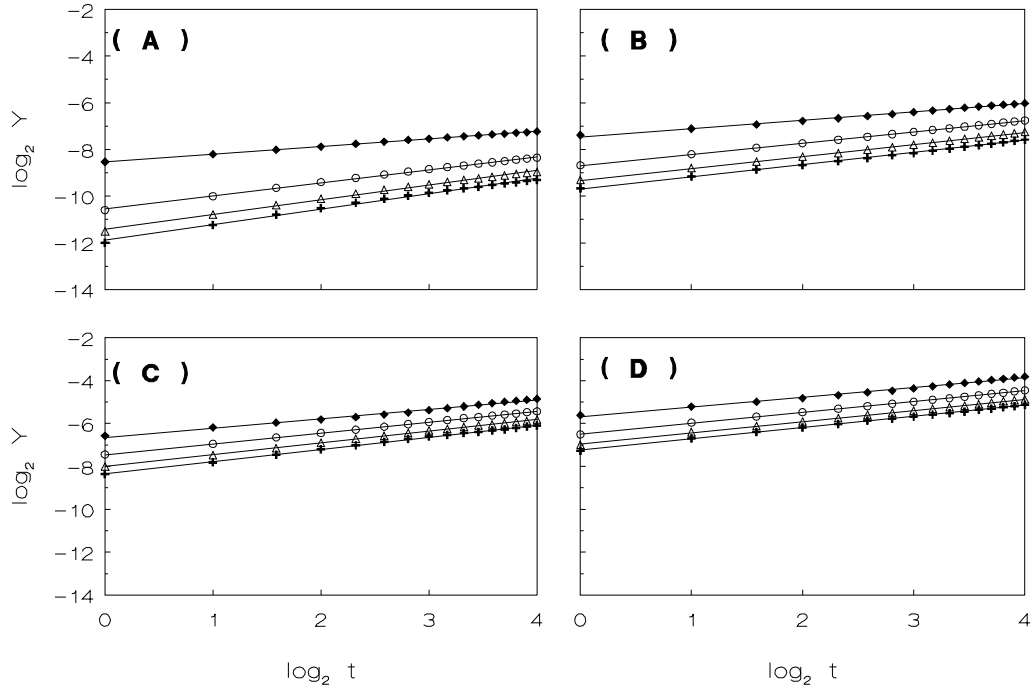}
}
\caption{Calculating generalized Hurst exponents for DAX index.
$q=0.5$, $1$, $2$, $4$ for crosses, triangles, circles,
and diamonds, respectively. Plots from A to D
are given for return times $\tau=1$, $10$, $60$~min. and
$1$~day. Solid lines represent least square linear fits.
}
\label{fig:Fig9}
\end{figure}

 The direct computation of the generalized Hurst exponents
was performed for return times $\tau = 1$, $10$ and $60$~min.
as well as for $\tau = 1$~day (as for the generalized Renyi
exponents of the graphs) and for the index
$q = 0.5$, $1.0$, $2.0$ and $4.0$.
The value $q=0$ cannot be used, as $q>0$
(see eq. (\ref{HurstDef})).
Computation of $H_q$ for the DAX index is displayed in Fig.~\ref{fig:Fig9}.
The linear scaling in the log-log plot is here excellent and
the corresponding results are collected in Table~\ref{Tab3}.
 One can clearly see that $H_q$ decreases with increasing $q$,
though this effect is smaller for longer time lags.
Also, for the smaller $q$-values it decreases with the increasing return
time $\tau$, at least within the range $\tau = 1\div 30$~min., and
the process is persistent ($H_1 > 1/2$). Very interestingly
however, for the larger $q$-values this tendency gets reversed:
$H_4(\tau)$ can be seen to approach the BRW limit from below
assuming values significantly lower than 1/2 at the high-frequencies.
For $\tau\simeq 10$~min. the exponent $H_1$ approaches the BRW value,
$1/2$.
 Notice, that at around $\tau_c \simeq 30$~min. the tail index is also
stabilized (see Sec.~II).
Similar critical value of $\tau$ was also reported for S\&P500
index in \cite{Stanley99} and estimated $\tau_c\simeq 20$~min.
(from (\ref{HurstDef}) it is clear that
the structure function $S_2(\tau)$ of the index
is equivalent to the volatility $v(\tau)$).
Hence, the correlations cease to exist slightly faster for DAX.
This may indicate that the correlations are weaker in the smaller market.
Also, for high frequency data the $q$--dependence is much stronger.
When $\tau$ grows, crossing the critical value $\tau_c$,
$H_q$ becomes almost independent of $q$.

For completeness we have performed the same computation for the DAX returns.
The quality of the linear scaling in this case is not as excellent
as for the index itself but still it quite consistently points to
$H_q\simeq 0.0$ for all the calculated values of $q$ and
$\tau$, the same result as for the WN (in contrast to the index that
considerably differs from BRW). This difference may be explained by
the existence of stronger correlations for the index.

\section{Discussion and summary}

 The financial processes are governed by a complex dynamics with
many degrees of freedom and various additional noise terms due to
complicated interactions with the external environment.
This is a non--stationary evolutionary process with strong correlations,
to large extent resembling biological evolution, and it cannot be viewed
as an equilibrium process \cite{Mccauley}.
From this perspective it is not very surprising that the
formalism that is so far available may not yet offer an optimal scheme
to consistently comprise all the related effects. The present
paper constitutes an attempt to provide some further relevant
empirical characteristics for another important world stock market,
the Deutsche Aktienindex, in addition to what already is available
in the literature. Several such characteristics, not always easy to
 interpret, are identified.

 In particular, the returns' pdf displays Pareto (fat) tails
of the type $1/x^\alpha$.
As was shown in Sec.~II the distribution of the DAX returns has its
central part close to the L\'evy distribution with $\alpha$
from $1.2$ (minutely returns) up to $\alpha \simeq 1.7$
(daily returns).
 Within the "tail region", {\it i.e.} for returns greater than
$3$ but below $10\div 20$ standard deviations,
a simple linear regression implies the tail index
$\alpha\simeq 2.4$ for minutely returns and up to
$\alpha\simeq 3.5$ for daily returns.
The available data sets seem to be too small to calculate
(with reasonable certainty level) the tail behavior beyond
$10\div 20$ standard deviations.
We find various tail
indices for various regions of the distribution, for different
return times and for different financial indices
(see {\it e.g.} \cite{Stanley99} for comparison).
 There is no doubt that the power like behavior plays important role,
but its manifestations are more complex than it was originally assumed.
It is worth to mention that similar multiscaling of pdf was found
in a monofractal toy model discussed by Bouchaud {\it et al}
\cite{Bouchaud}.

 From the dynamical point of view more important than pdf are Hurst
and Renyi
exponents of the index and returns' graph. This is because they
take into account the time dependence of the process.
Therefore, the main part of our analysis was devoted to study
these quantities.
 For regular (mono--) fractal set the following simple relation
between fractal exponent of the graph ($d_0$) and the Hurst exponent
is often satisfied \cite{Mandel,Skjeltrop,Ausloos01,Jaffard97}
\begin{equation}
d_0 = 1 + E - H_1
\label{dHrel}
\end{equation}
where $E$ is the embedding dimension of the data series
($E=1$ for a one dimensional data).
 In particular, this relation is preserved for the WN,
for the BRW as well as for the Weierstrass--Mandelbrot
fractal \cite{WMfractal} and for any fractional Brownian motion
\cite{Mandel}.
This relation is usually used when discussing the
fractal behavior of financial indices.
However, the often cited relation (\ref{dHrel}) is
not fulfilled in our case. In general, the multifractal
formalism for a function, instead of the exact spectrum,
yields an upper bound of its H\"older spectrum
\cite{Mandel,Jaffard97}.
 Hence, in principle it is necessary to calculate the Renyi exponents
by direct application of the box counting algorithm for the graphs.

 We have used the (2--dimensional) box counting
algorithm to calculate the fractal properties of the financial
graphs. The Renyi exponents with $q=0.5$, $1.0$, $2.0$
and $4.0$ for the DAX index were found to be around $1.3$,
slowly decreasing with growing $q$ (about $15$\% below
the constant BRW value, $1.5$).
However, for the DAX returns the corresponding values turn out
not to be so precisely determined due to a somewhat poorer scaling
but still can be localized within the range $1.4\div 1.7$,
more below the WN value and with stronger $q$--dependence.
The decrease of $d_q$ with growing $q$, usually viewed as
a sign of multifractality, is bigger here than for the index itself,
especially for the high--frequency data (1~min. returns).
 It is worth to notice that the estimated difference between
 the fractal exponents of the index and exponents of the corresponding
 differential series (returns) is only about $0.2\div 0.3$,
while it is exactly equal to $0.5$ for the BRW and WN.
As a potential most visible and significant difference
between the behavior of the financial index and BRW, this effect
demands however further more systematic study.
 Comparing the above results with the calculated Hurst exponents
one can also see that the relation (\ref{dHrel}) is violated
for the index and seems to be violated even more for the returns.
 In contrast to the Renyi exponents, the generalized Hurst exponents
for returns are compatible with those of the WN (equal to zero).
For the index we obtain persistent behavior ($H_1= 0.51\div0.63$)
which is slightly stronger for short time lags.
Here, the $q$--dependence is also stronger than for the $d_q$.

 One can conclude that the DAX data series, both for the
(logarithm of) index and its returns, have very complicated
self--similar structures that may escape any unique multifractal description.
 One likely reason is that the real financial data
series are superpositions of series with different
properties, {\it e.g.} for the "regular" and "rush" hours
as was suggested recently \cite{Huang00,Stan1,Stan2,Hupp}.
To clarify these intriguing effects further empirical and theoretical
analysis is necessary.

\acknowledgments
We thank F. Gr\"ummer and J. Kwapie\'n for helpful exchanges.
M.~Frame and J.~Szmigielski of Yale Math. Dept. are acknowledged
for clarifying comments concerning fractal and Hurst exponents.

\begin{table}
\caption{Renyi box counting exponents ($d_q$) for DAX index}
\label{Tab1}
\begin{tabular}{c||cccc}
$\tau$&$q=0.5$&$q=1$&$q=2$&$q=4$\\
\tableline\tableline
 1~min.&1.38&1.36&1.34&1.30\\
10~min.&1.36&1.35&1.33&1.29\\
60~min.&1.30&1.30&1.29&1.26\\
  1~day&1.33&1.33&1.32&1.29\\
\end{tabular}
\end{table}

\begin{table}
\caption{Hurst exponents ($H_q$) for DAX index}
\label{Tab3}
\begin{tabular}{c||cccc}
$\tau$&$q=0.5$&$q=1$&$q=2$&$q=4$\\
\tableline\tableline
 1~min.&0.66&0.63&0.56&0.33\\
10~min.&0.52&0.51&0.48&0.36\\
60~min.&0.57&0.55&0.51&0.44\\
  1~day&0.52&0.52&0.51&0.46\\
\end{tabular}
\end{table}


\end{document}